\RequirePackage[hyphens]{url} % avoids overflow URL
\documentclass[12pt,a4paper,fleqn]{article}

\usepackage[textheight=23cm,textwidth=15cm]{geometry}
\usepackage{amsmath}
\usepackage{graphicx}
\usepackage{xcolor}
\usepackage{url}
\usepackage[american]{babel}
\usepackage{here}
\usepackage[articletitle=true, super=false, maxauthors=500]{achemso}
\setkeys{acs}{biblabel = brackets}

\usepackage{siunitx}
\PassOptionsToPackage{hyphens}{url}\usepackage{hyperref}
\hypersetup{colorlinks,linkcolor=black,urlcolor=blue, citecolor=black}

\begin{document}

\begin{center}

{\LARGE\bf
{\sc Heron}: Visualizing and Controlling Chemical Reaction Explorations and Networks
}

\vspace{1cm}
{\large
Charlotte H. M\"uller$^{\ddagger, }$\footnote{ORCID: 0000-0002-6640-5065}
, Miguel Steiner$^{\ddagger, }$\footnote{ORCID: 0000-0002-7634-7268}
, Jan P. Unsleber$^{\ddagger, }$\footnote{ORCID: 0000-0003-3465-5788}
, Thomas Weymuth\footnote{ORCID: 0000-0001-7102-7022}
, Moritz Bensberg\footnote{ORCID: 0000-0002-3479-4772}
, Katja-Sophia Csizi\footnote{ORCID: 0000-0001-8352-1060}
, Maximilian M{\"o}rchen\footnote{ORCID: 0000-0002-7467-5719}
, Paul L. T\"urtscher\footnote{ORCID: 0000-0002-7021-5643}
\:and Markus Reiher$^{\ast, }$\footnote{Corresponding author; e-mail: mreiher@ethz.ch; ORCID: 0000-0002-9508-1565}
}\\[4ex]

ETH Zurich, Department of Chemistry and Applied Biosciences,\\
Vladimir-Prelog-Weg 2, 8093 Zurich, Switzerland.\\
$^{\ddagger}$ Authors contributed equally\\

October 07, 2024

\vspace{.43cm}

\textbf{Abstract}
\end{center}
\vspace*{-.41cm}
{\small
Automated and high-throughput quantum chemical investigations into chemical processes have become feasible in great detail and broad scope.
This results in an increase in complexity of the tasks and in the amount of generated data.
An efficient and intuitive way for an operator to interact with these data and to steer virtual experiments is required.
Here, we introduce \textsc{Heron}, a graphical user interface that
allows for advanced human-machine interactions with quantum chemical exploration campaigns into molecular structure and reactivity.
\textsc{Heron} offers access to interactive and automated explorations of chemical reactions with standard electronic structure modules, haptic force feedback, microkinetic modeling, and refinement of data by automated correlated calculations including black-box complete active space calculations.
It is tailored to the exploration and analysis of vast chemical reaction networks.
We show how interoperable modules enable advanced workflows and pave the way for routine low-entrance-barrier access to advanced modeling techniques.
}

\newpage

\section{Introduction}
\label{sec:introduction} 
Computational chemistry provides a detailed description of chemical processes, ranging from predicted spectra of compounds to the exploration and characterization of their chemical reactivity.
Due to hardware and software advances, the automated first-principles exploration of reaction mechanisms~\cite{Vazquez2018, Dewyer2018, Simm2019, Unsleber2020, Maeda2021, Baiardi2022, Steiner2022, Ismail2022}, a field in which elementary transformations are elucidated to build a network of all chemical compounds and their reactions in a chemical system, has emerged to replace slow and limited manual computations.
The resulting chemical reaction networks allow for understanding chemical processes in great detail.
Uncovered side reactions and alternative pathways within the network are key for improving an investigated process or designing new reactive systems with specific properties.
However, the generated amounts of data, linking individual structures with elementary transformations and annotating these structures with molecular properties, can quickly become challenging to manage.~\cite{Steiner2022}
The inspection of a reaction network, both in terms of its analysis and the steering of an exploration, is therefore a nontrivial task.
Hence, dedicated software is required to promote easy interaction with automatically explored reaction networks by suitable visualization of task-focused operations and data selection.

For this purpose, we introduce \textsc{Heron} as a graphical user interface (GUI) for our \textsc{Scine} (Software for Chemical Interaction Networks)~\cite{Scine} software project.
\textsc{Heron} is open source and free of charge.
In addition to the publicly available releases~\cite{Heron2022}, \textsc{Heron} is also part of the AutoRXN workflow~\cite{Unsleber2023} on Microsoft Azure and Microsoft Quantum Elements~\cite{QuantumElements, QuantumElementsEvent}.
Furthermore, \textsc{Heron} features a graphical integration of the \textsc{autoCAS}~\cite{Stein2016, Stein2016b, Stein2017, Stein2017b, Stein2019} software that allows for selecting active orbital spaces in an automated fashion and then automatically carrying out complete active space (CAS) calculations.
Moreover, \textsc{Heron} provides interactive real-time quantum mechanics with force feedback which can be applied in research~\cite{Haag2011, Haag2013, Haag2014, Vaucher2016, Vaucher2017} and in educational settings~\cite{Weymuth2021, Muller2023a}.
\textsc{Heron} adopts a modular approach where its functionality is encapsulated into different views. 
For example, the interactive manipulation of a molecule is one module and viewing the results of a rolling automated reaction-mechanism exploration by the software \textsc{Chemoton}~\cite{Unsleber2022, Chemoton2023} is another. In addition to the documentation included in every software release, a video tutorial of the most important functionalities is available at \url{https://youtu.be/HGXkir9iKFo}.

Different software packages dedicated to visualizing and analyzing total reaction networks have already been reported~\cite{Cheng2014, Gupta2020, Kuwahara2023}, and additional schemes based on existing graph visualization software~\cite{McDermott2021, acerxn} or manual layouts are available.
However, the \textit{entire} network rarely needs to be inspected to understand a chemical process.
Hence, \textsc{Heron} focuses its network views on subsections and paths, allowing for a visualization that remains uncluttered for all but the largest networks where already all favorable reactions of a single compound generate a
graph that is difficult to comprehend.

This work is organized as follows:
We discuss structure preparation, running quantum chemical calculations, integration of haptic real-time quantum chemistry, and controlling a rolling reaction network exploration with \textsc{Heron} in sections~\ref{sec:interactive} and \ref{sec:chemoton}. We then detail how \textsc{Heron} allows the operator to inspect large network databases, visualize chemical reaction networks, identify paths in large reaction networks, and incorporate information from microkinetic modeling in sections~\ref{sec:pathfinder} and \ref{sec:kinetic_modelling}. This is followed by an overview of correlated calculations and the automated selection of active orbital spaces through \textsc{Heron} in section~\ref{sec:autocas} and the construction of interactive hybrid models in section~\ref{sec:interactive_qmmm}. A technical description of its software design is provided in appendix~\ref{app:design}.

\section{Methods}

\begin{figure}[H]
    \centering
    \includegraphics[width=0.9\textwidth]{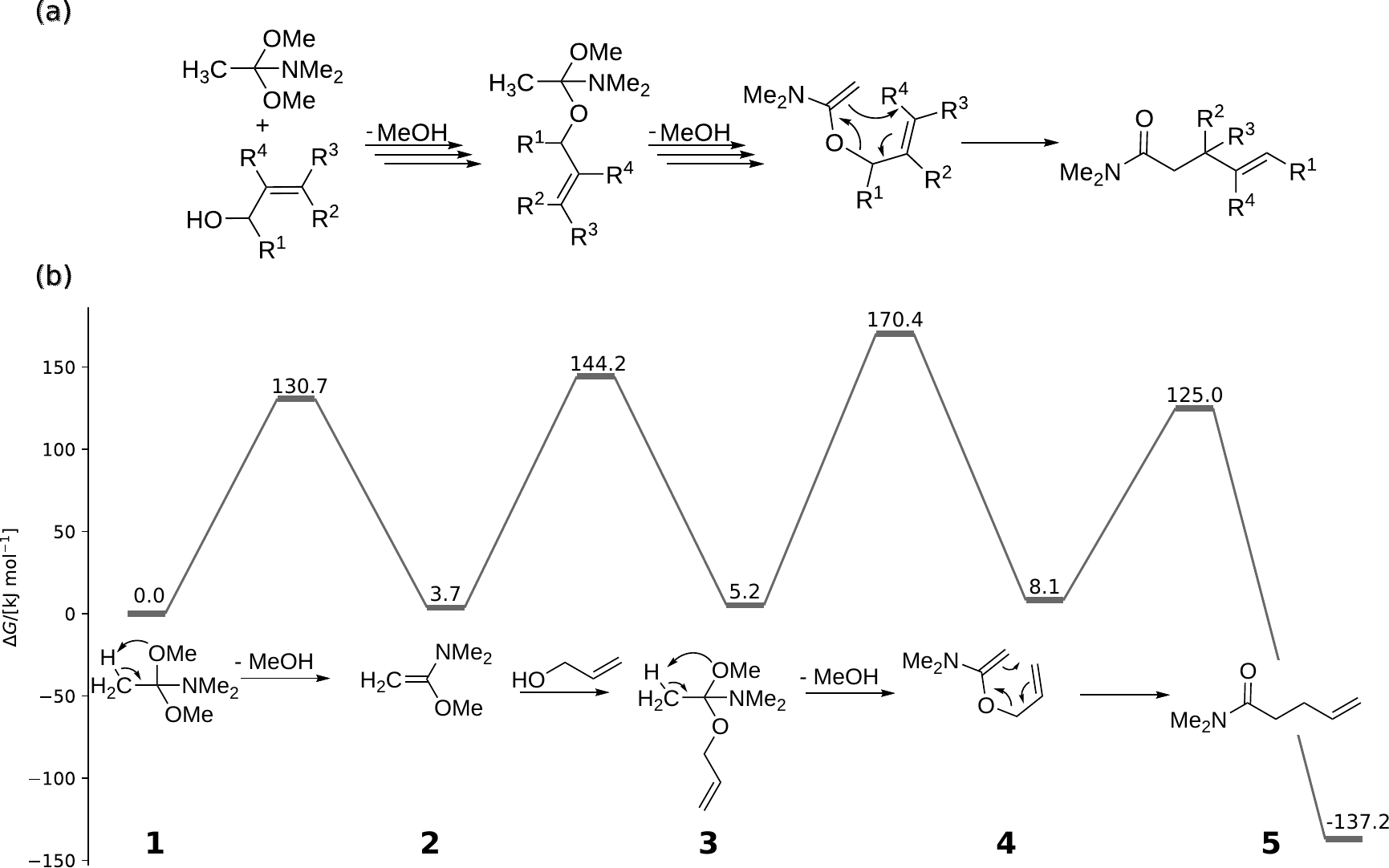}
    \caption{(a) Lewis structures of the Eschenmoser--Claisen rearrangement. The multiple arrows of the first and second step highlight that these are not necessarily elementary reaction steps. (b) Free energy diagram and the reaction mechanism found by \textsc{Chemoton} in a fully automated exploration taken from Ref.~\cite{Bensberg2024a}. All energies in $\si{kJ.mol^{-1}}$ calculated with PBE0-D3BJ(Toluene)//GFN2-xTB(Toluene).}
    \label{fig:eschenmoser_claisen_mechanism}
\end{figure}

We demonstrate the features of \textsc{Heron} with the example of the Eschenmoser--Claisen rearrangement, which we have recently studied in a kinetically steered first-principles exploration~\cite{Bensberg2024a}.
In 1964, Eschenmoser and coworkers discovered that allylic and benzylic alcohols undergo a Claisen rearrangement if heated with N,N-dimethylacetamide dimethyl acetal in xylenes, forming a $\gamma$,$\delta$-unsaturated amide as the product~\cite{Wick1964}. 
Because this Eschenmoser--Claisen rearrangement is more $(E)$-selective in the formation of 
the double bond and takes place at lower temperatures ($100-150~\si{\celsius}$) than other Claisen rearrangements, it has been used in the total synthesis of natural products~\cite{Daniewski1992, Chen1993, Williams2000, Loh2000}.
The key steps of the reaction mechanism of the Eschenmoser--Claisen rearrangement are shown in Figure~\ref{fig:eschenmoser_claisen_mechanism}(a).
In the first step, the unsaturated alcohol exchanges one of the methoxy groups of N, N-dimethylacetamide dimethyl acetal. 
The elimination of the second methoxy group and deprotonation of the methyl group follows. Then, the Claisen rearrangement occurs, forming the final product. For an illustrative model system of the reaction, we replaced all side chains R$^\mathrm{X}$ (X = 1, 2, 3, 4), shown in Figure~\ref{fig:eschenmoser_claisen_mechanism}(a), by hydrogen atoms and started the exploration from N,N-dimethylacetamide dimethyl acetal and allyl alcohol. 
We extracted a reaction mechanism with \textsc{Heron} from the PBE0-D3BJ//GFN2-xTB reaction network of Ref.~\cite{Bensberg2024a} by considering the concentration fluxes along the reactions based on the visualization presented in section~\ref{sec:kinetic_modelling} below.
The mechanism and the relative free energies encoded in the network are shown in Figure~\ref{fig:eschenmoser_claisen_mechanism}(b). In the first two steps of our mechanism, methanol and allyl alcohol are exchanged at the starting acetal \textbf{1}. Because no proton source is available in toluene, the enamine \textbf{2} is formed as an intermediate. The slowest step in the reaction mechanism (with a relative activation free energy $\Delta G^\ddagger = 182.3~\si{kJ.mol^{-1}}$) is the dissociation of the second methanol molecule to form the 1,2 and 5,6 unsaturated Claisen rearrangement precursor \textbf{4}. The [3,3] sigmatropic rearrangement shows a significantly lower barrier than the preceding methanol dissociation of $\Delta G^\ddagger = 124.8~\si{kJ.mol^{-1}}$ compared to the starting reactants. We found in Ref.~\cite{Bensberg2024a}
the reaction to be highly exergonic with a reaction free energy of $\Delta G = -125.9~\si{kJ.mol^{-1}}$.

\section{Results and Discussion}

\subsection{Structure Preparation, Manipulation, and Haptic Real-Time Quantum Chemistry}
\label{sec:interactive}
\begin{figure}[H]
    \centering
    \includegraphics[width=\textwidth]{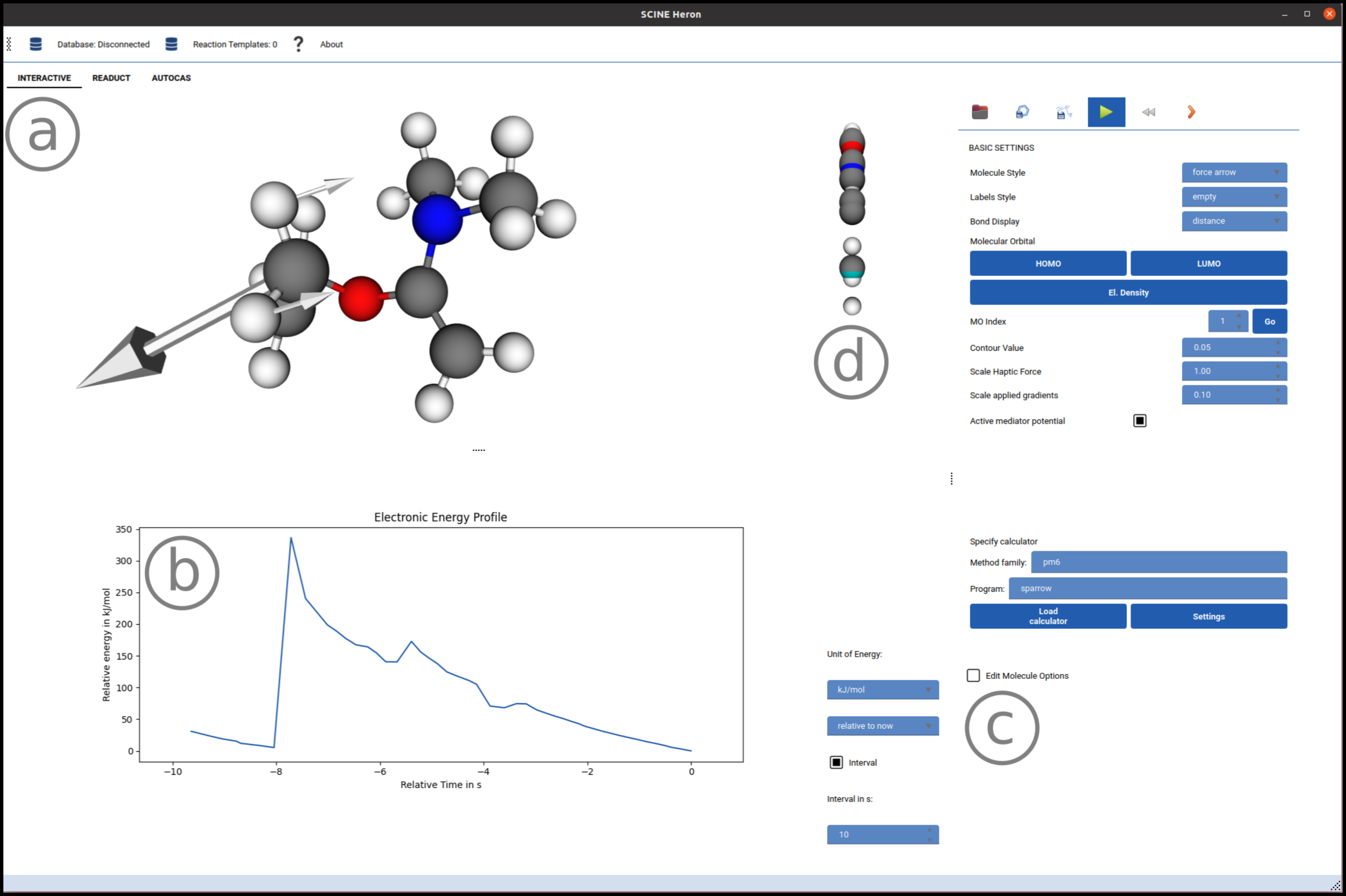}
    \caption{
    The interactive module in \textsc{Heron} consists of a molecular viewer (a), the display of a total electronic energy graph (b), and the display of settings (c). The spatial challenge of translating three-dimensional movement to the two-dimensional screen is facilitated by a visual representation of the atoms projected onto the axis perpendicular to the screen (d). The forces acting on individual atoms can be displayed as arrows. The force acting on a selected atom can be experienced as repulsion or attraction through a haptic force-feedback hardware device.
    Oxygen atoms are colored red, carbon atoms gray, nitrogen atoms blue, and hydrogen atoms white. The turquoise sphere in (d) represents the current depth of the haptic device pointer.
    }
    \label{fig:interactive}
\end{figure}

A molecular structure can be created directly within the environment (by placing individual atoms), loaded from file (supported formats are XYZ, PDB, or MOL) or transferred from one of the other tabs of \textsc{Heron}. Interactive quantum mechanics~\cite{Haag2013} allows for intuitive and efficient exploration of potential energy hypersurfaces and, thereby, promotes the discovery of potentially unexpected intermediates and transition state structures~\cite{Haag2011, Haag2014}.
A module combines the frameworks of real-time quantum chemistry~\cite{Haag2013} and haptic quantum chemistry~\cite{Marti2009} in a visual and haptic interface.
The GUI design consists of a molecular viewer with a corresponding total electronic energy graph and multiple calculation and visualization settings (see Figure~\ref{fig:interactive}).

The operator can then interact with the molecular structure by pulling on one or multiple atoms, while, at the same time, the structure is continuously optimized in real time.
The required visual feedback of 60~Hz and haptic feedback of 1 kHz is achieved by efficient quantum chemical methods~\cite{Bosia2023}. Available for this purpose in \textsc{Heron} are density functional tight binding methods~\cite{Porezag1995, Seifert1996, Elstner1998, Gaus2011}, methods which neglect diatomic differential overlap~\cite{Dewar1977, Thiel1992, Dewar1985, Rocha2006, Stewart1989a, Stewart1989b, Stewart2007}, and extended tight binding methods~\cite{Bannwarth2020} through the \textsc{Scine} package \textsc{Sparrow}~\cite{Bosia2023, Sparrow2023}.
We note that interactive quantum mechanics has even been demonstrated for full-fledged density functional theory methods~\cite{Luehr2015}.
To achieve smooth operability, especially in the case of slow orbital optimization convergence, a mediator potential approximating the energy and gradients by a second-order Taylor expansion at the coordinates of the last calculated structure is available~\cite{Vaucher2016}.

Structure manipulation can either be carried out with a computer mouse or with a $\mathrm{Touch}^\mathrm{TM}$ haptic device.
The latter communicates with the graphical user interface through the open-source \textsc{OpenHaptics} library~\cite{Itkowitz2005}.
While the computer mouse is more familiar and readily available, the haptic device provides haptic feedback on the energy gradient along the current direction of motion.
Such a setting delivers an intuitive understanding of a molecular system, which can be advantageous for research~\cite{Vaucher2017} and learning~\cite{Weymuth2021, Muller2023a}. For example, the Eschenmoser--Claisen rearrangement can be easily reproduced interactively by manipulating the atomic positions in real time while also receiving sensory feedback on the activation barrier (see Figure~\ref{fig:interactive}).

The settings are divided into basic settings that are intuitively understandable (molecular charge, mode of representation) and advanced settings that may only be needed in specific cases (methods and editing). 

\begin{figure}[H]
    \centering
    \includegraphics[width=\textwidth]{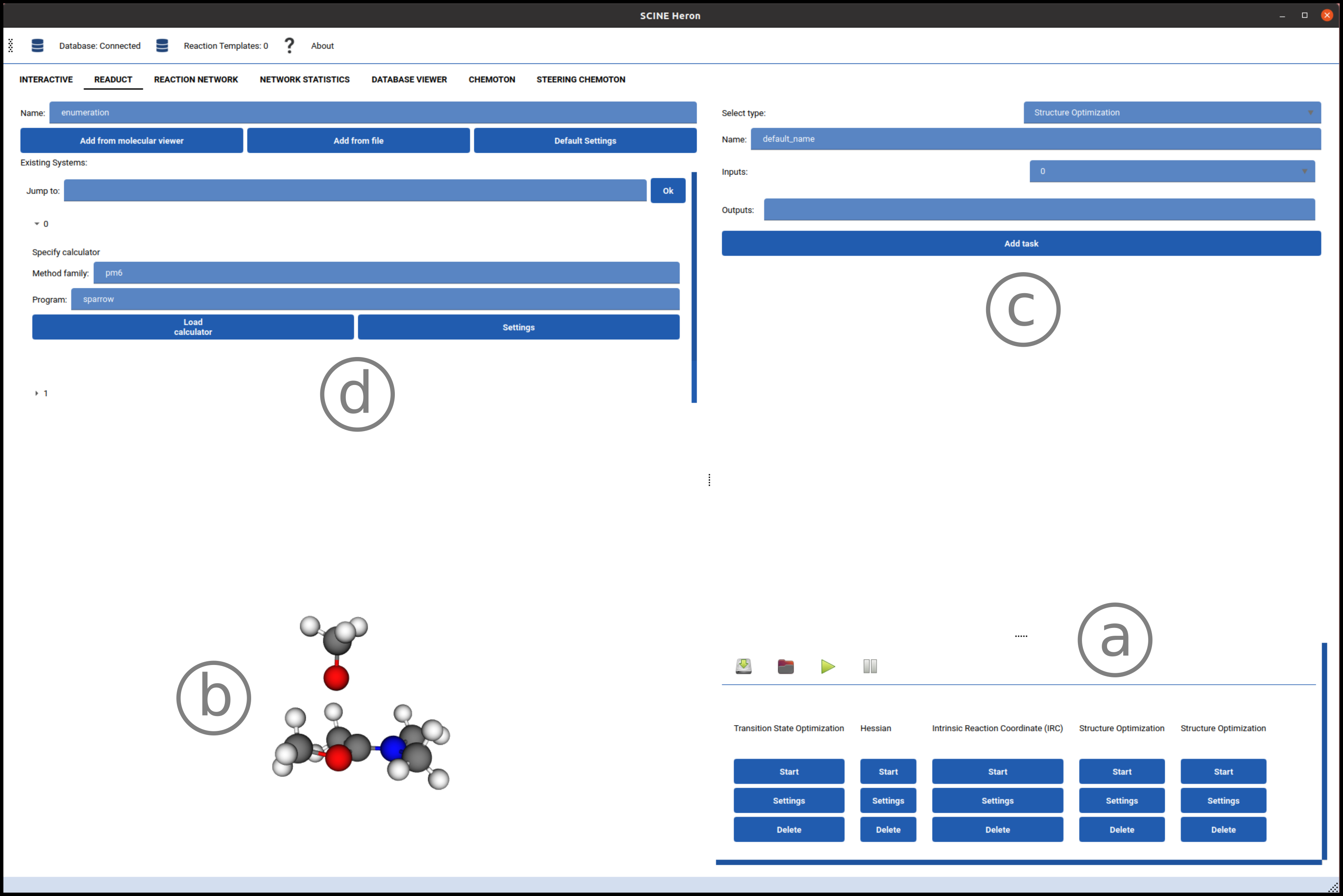}
    \caption{The \textsc{ReaDuct} interface within \textsc{Heron}, featuring a pipeline of different standard structure optimization algorithms (a) for a given structure (b). New tasks can be constructed through a pull-down menu (c), and all optimized and manually added structures are stored in random access memory (d).}
    \label{fig:readuct}
\end{figure}

For quantitative studies, structures and paths can be directly transferred to stationary point or minimum energy path search algorithms implemented in \textsc{ReaDuct}~\cite{Vaucher2018, Readuct2023}.
In \textsc{Heron}, multiple structures can be stored and subjected to minimization and transition state searches by selection from a pull-down menu.
Expert settings for the electronic structure calculations and for the optimization routines can be adjusted in separate pop-up windows.
Additionally, more complex optimization workflows (such as structure optimization of reactant and product, then double-ended transition state search, followed by transition state optimization by partitioned rational function optimization~\cite{Banerjee1985}, and subsequent intrinsic reaction coordinate scans~\cite{Fukui1970}) can be constructed by straightforward task chaining (see Figure~\ref{fig:readuct}). 
These workflows can also be imported from and exported to human-readable YAML-format files, which can then be run on the command line, allowing a seamless switch between \textsc{Heron} and the command line interface to \textsc{ReaDuct}.

\subsection{Chemical Reaction Networks}
\label{sec:chemoton}

\subsubsection{First-principles Exploration}
\label{subsec:exploration}

Within the \textsc{Scine} framework~\cite{Scine}, fully automated reaction network explorations can be carried out with \textsc{Chemoton}~\cite{Unsleber2022, Chemoton2023}, which is a flexible and general reaction-mechanism exploration program. It offers numerous options to customize and steer autonomous explorations.
However, extensive fine-tuning of exploration options, especially for exploration strategies that involve multiple sequential steps, can require expert knowledge and cautious oversight over the exploration progress.
To address this issue, \textsc{Heron} simplifies the interaction with \textsc{Chemoton}.

\begin{figure}[H]
	\centering
	\includegraphics[width=\textwidth]{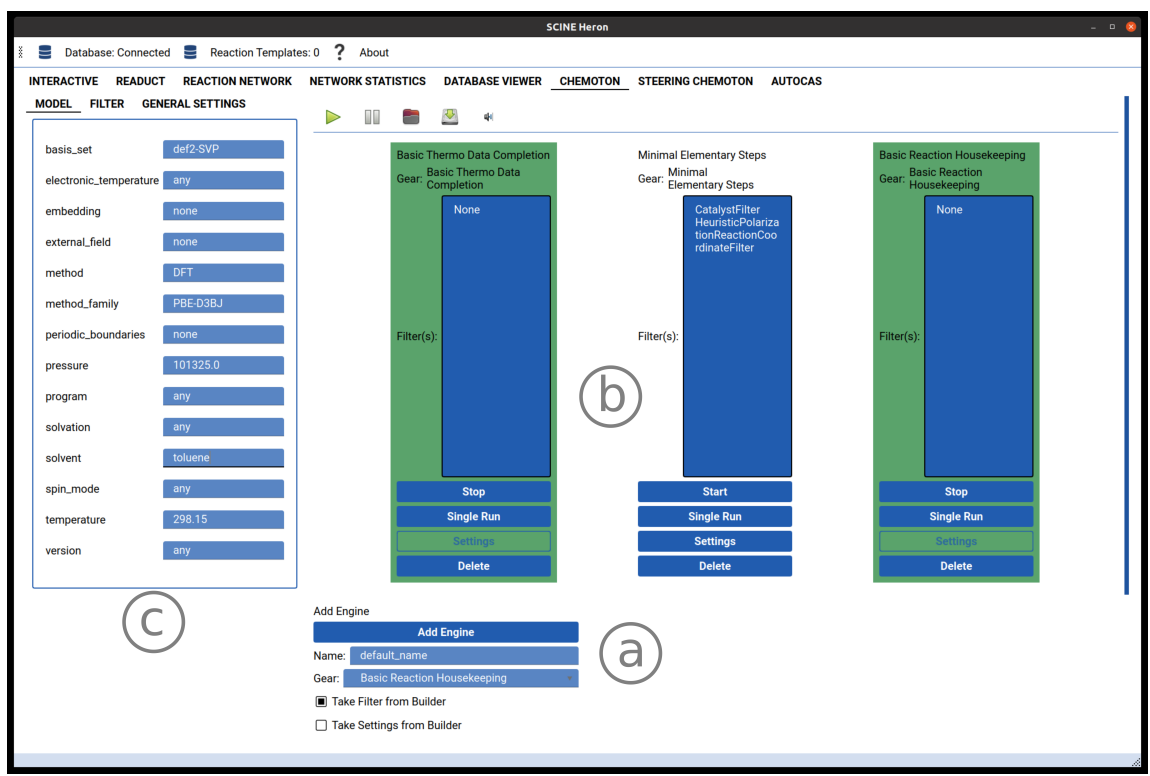}
    \caption{\label{fig:chemoton}
    \textsc{Heron} tab that allows one to execute individual \textsc{Chemoton} \texttt{Engines}; (a) Menu to add a new \texttt{Engine} to the exploration; (b) Overview of the current exploration (green background marks that the \texttt{Engine} is currently running). (c) Menu to define global exploration settings.}
\end{figure}

\textsc{Heron} enables complete control of all, otherwise autonomous algorithms in \textsc{Chemoton}.
The corresponding interface is shown in Figure~\ref{fig:chemoton}.
It leverages \textsc{Chemoton}'s design principle of non-interacting microservices, called \texttt{Engines}.
Every exploration with \textsc{Chemoton} is defined by a set of \texttt{Engines} that fulfill a specific task.
Such tasks include reactivity assessment, calculation setup, and collection of results.
\textsc{Heron} lets the operator build an exploration campaign by selecting individual \texttt{Engines} from a pull-down menu (Figure~\ref{fig:chemoton}(a)).
For the Eschenmoser--Claisen exploration, one can, for example, select a \texttt{BasicThermoDataCompletion Engine} and run it indefinitely to calculate a Hessian for all explored minimum energy structures and assign a Gibbs free energy correction to them.
The selected \texttt{Engines} are added as boxes to the overview panel (Figure~\ref{fig:chemoton}(b)), where they can be customized and switched on and off, signaled by their green or transparent background.
The options of each \texttt{Engine} can be edited individually.
To avoid specifying the same options repeatedly, globally applicable options, such as the electronic structure model, can be edited beforehand on the left-hand side (Figure~\ref{fig:chemoton}(c)).
To avoid repetitive tasks further, \textsc{Heron} offers a binary and human-readable JSON format for the exploration protocol.
This allows one to share and distribute exploration protocols and common setups.

\subsubsection{Activating Filters for Selective Explorations}
\label{subsec:filters}

\begin{figure}[H]
	\centering
	\includegraphics[width=0.5\textwidth]{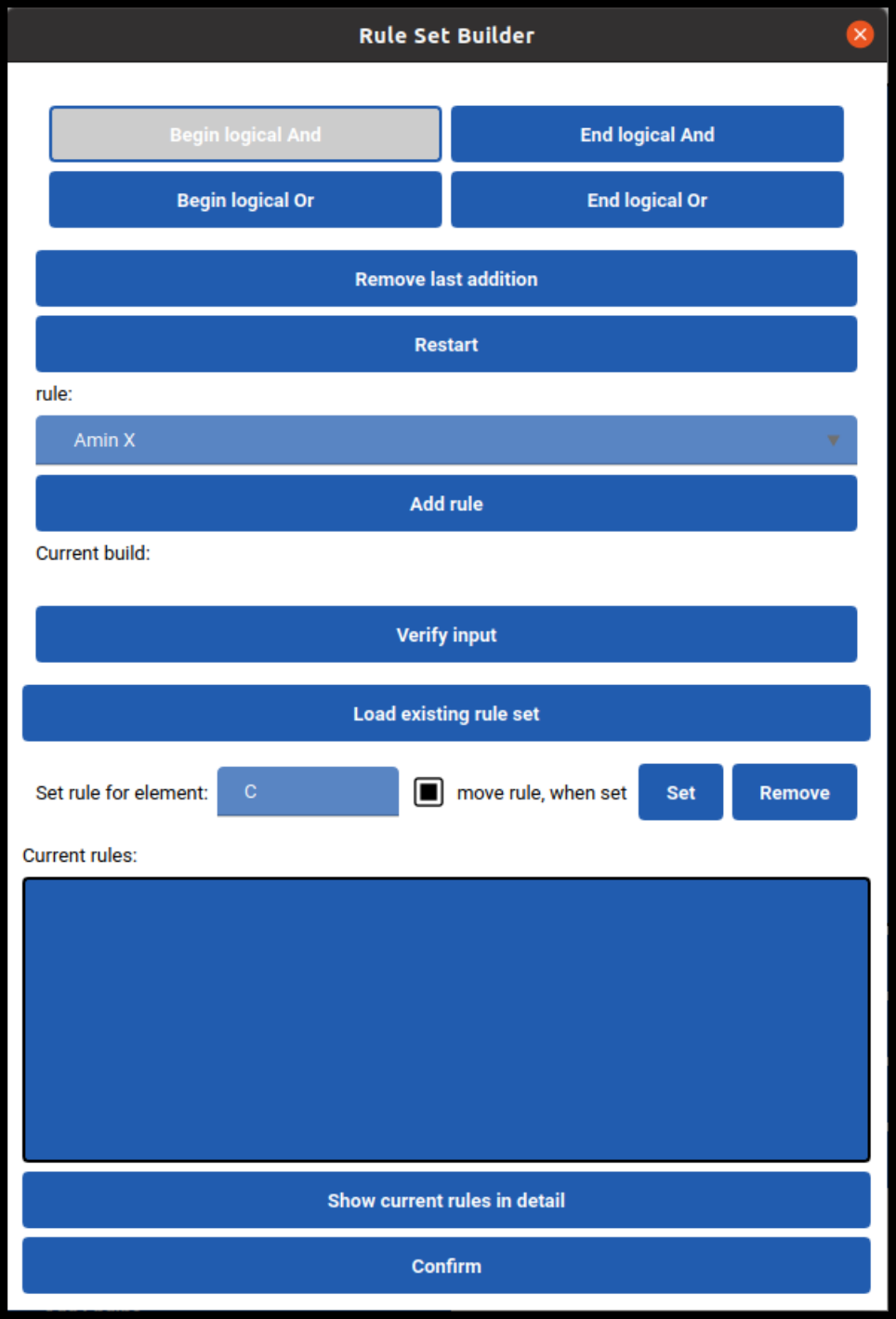}
    \caption{\label{fig:rules-builder} The menu 
    in \textsc{Heron} to build a set of reaction rules. All reactivity filters of \textsc{Chemoton}, including reactive site filters that require graph-based or electronegativity-based rules, can be built. Various rules can be combined by logical operations \textit{'and'/'or'}, allowing a high degree of specificity.}
\end{figure}

Since \textsc{Chemoton} considers every compound as reactive with every atom as a potentially reactive site, a large number of potential reactions arise from the combinatorial possibilities. 
To avoid a combinatorial explosion of the required reaction trials in an autonomous exploration, it is advisable to introduce heuristic values to limit the search space. 
In \textsc{Chemoton}, this concept has been introduced as so-called filters~\cite{Unsleber2022}.
All filters implemented in \textsc{Chemoton} can be constructed directly in \textsc{Heron} as shown in Figure~\ref{fig:rules-builder}.
A built set of reaction rules can be stored in either a binary or human-readable JSON format.
This feature facilitates the creation of a library of fundamental rules for straightforward guidance of explorations, the usability of \textsc{Chemoton}, and the reproducibility of studies carried out within \textsc{Heron}.
We provide a set of graph-based rules for organic chemistry in the accompanying release to this paper.

\subsubsection{Driving an Exploration with a Steering Wheel}
\label{subsec:steering}

\begin{figure}[H]
	\centering
	\includegraphics[width=1.0\textwidth]{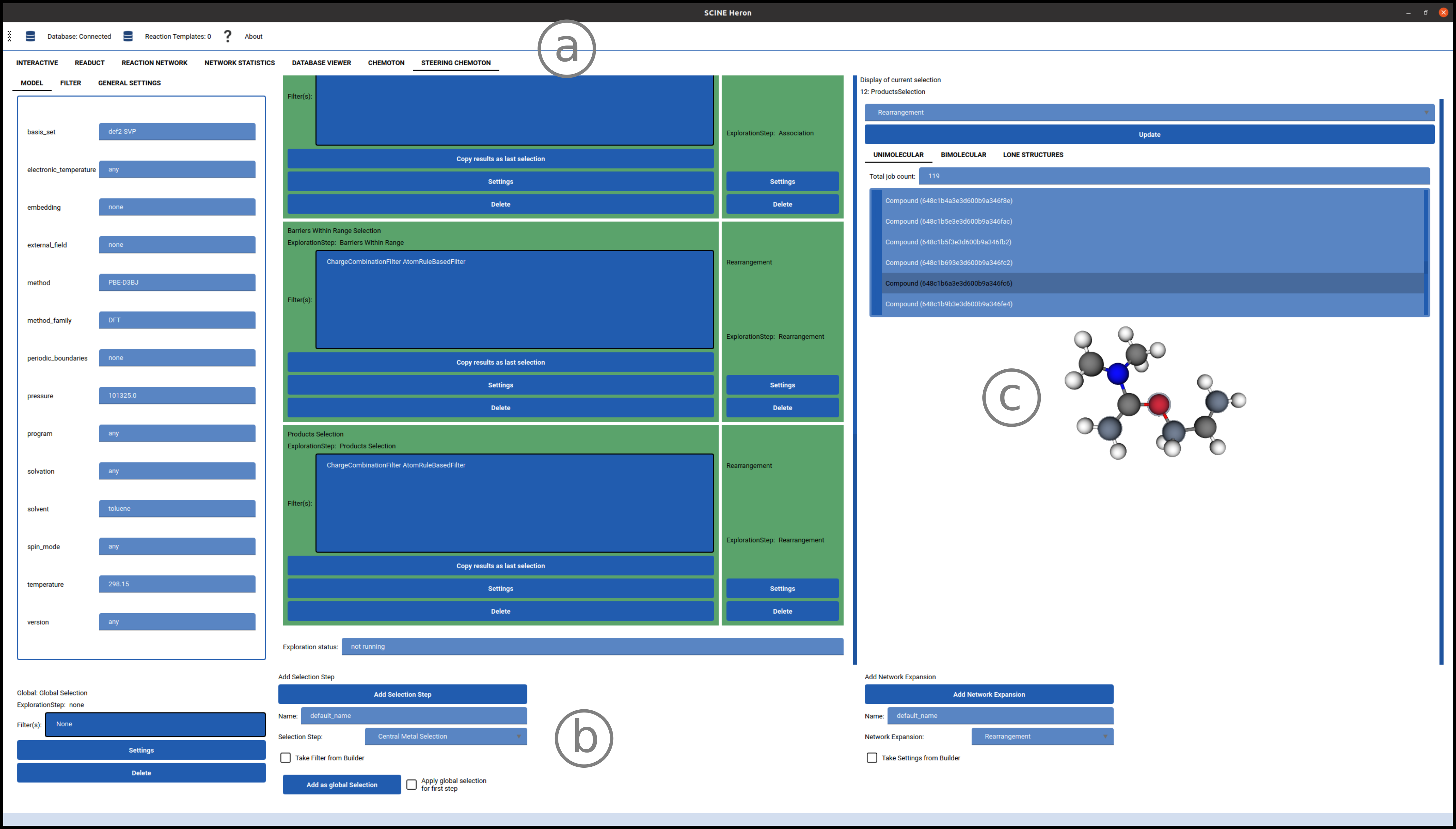}
    \caption{\label{fig:steering} The \textsc{Steering} \textsc{Wheel} in \textsc{Heron}. (a) An active exploration protocol. The green background color marks successful execution. (b) Two pull-down menus allow one to add a \texttt{Selection Step} or \texttt{Network Expansion Step} to the protocol; (c) The right console allows one to query the latest \texttt{Selection Step} for a potential next \texttt{Network Expansion Step} (here, a dissociation reaction was selected as a potential next \texttt{Network Expansion}, so that the selected subset of the reaction network is probed for dissociation reactions). All resulting calculations that will be set up are displayed in the table with a unique identifier.
    Each potential reactive complex can be visualized. The blue transparent spheres mark the reactive sites of the structure shown.}
\end{figure}

If applied to specific chemical problems, such as catalyst design and other chemical optimization challenges, a more directed approach can be advantageous because widespread autonomous approaches can be stalled by the combinatorial explosion of reaction trials (even when filters are applied). Then, computing time would be wasted on regions of the chemical reaction space that might not be of immediate interest to the operator.
A directed approach should strike a balance between flexibility, coverage of the chemical space, extensibility to autonomous explorations, and reproducibility of the executed protocol.
We achieved all of this by replacing the selection of individual \texttt{Engines} with the selection of exploration steps that build an exploration protocol called \textsc{Steering Wheel}~\cite{Steiner2023}.
The \textsc{Steering Wheel} allows one to drive actively autonomous explorations through chemical reaction space as shown in Figure~\ref{fig:steering}.

The \textsc{Steering Wheel} consists of two alternating exploration control classes: \texttt{Network Expansion} and \texttt{Selection Step} (see Ref.~\citenum{Steiner2023} for details).
These two control categories allow one to navigate the explored chemical space step by step.
The exploration protocol, shown in Figure~\ref{fig:steering}(a), can be built within \textsc{Heron} by choosing steps from pull-down menus, shown in Figure~\ref{fig:steering}(b).
At the beginning of an exploration, it may be unclear how to construct the full protocol because the best possible option for later steps in the exploration will depend on the previous results. 
Therefore, it must be easily understandable what the current exploration status is and how planned future steps will potentially affect the exploration.
The steering tab in \textsc{Heron} solves this problem by displaying how a potential next \texttt{Network Expansion Step} would affect the exploration, shown in Figure~\ref{fig:steering}(c).
The operator can select a potential next \texttt{Network Expansion Step}, then the number of calculations set up by such an expansion, alongside the constructed reactive complexes and their reactive sites, are displayed based on a database query, which enables one to refine the chosen \texttt{Network Expansion} or \texttt{Selection Step} according to the targeted chemical space.
To ensure that the exploration process, including the taken steps that define the explored chemical space,
can be reproduced and shared with others, any exploration protocol can be saved in a binary and human-readable JSON file format.

One such exploration protocol is provided on Zenodo together with the resulting database~\cite{HeronZenodoSI2024}.
It encodes a steered exploration for the Eschenmoser--Claisen rearrangement based on the expected mechanism of S\textsubscript{N}1 substitution of methanol with the allyl alcohol, second methanol elimination, and final electrocyclic rearrangement.
It can reproduce the results of the kinetically-guided exploration in section~\ref{sec:kinetic_modelling}
while lowering the required reaction trials from 19,400 to 12,600 and increasing the number of found nodes corresponding to unique molecular aggregates (compounds and flasks) from 2,220 to 2,411.
The steered exploration was set up and executed completely within \textsc{Heron} and required seven \texttt{Network Expansion Steps}.

\subsubsection{Reaction Templates}
\label{subsec:templates}
\begin{figure}[H]
    \centering
    \includegraphics[width=\textwidth]{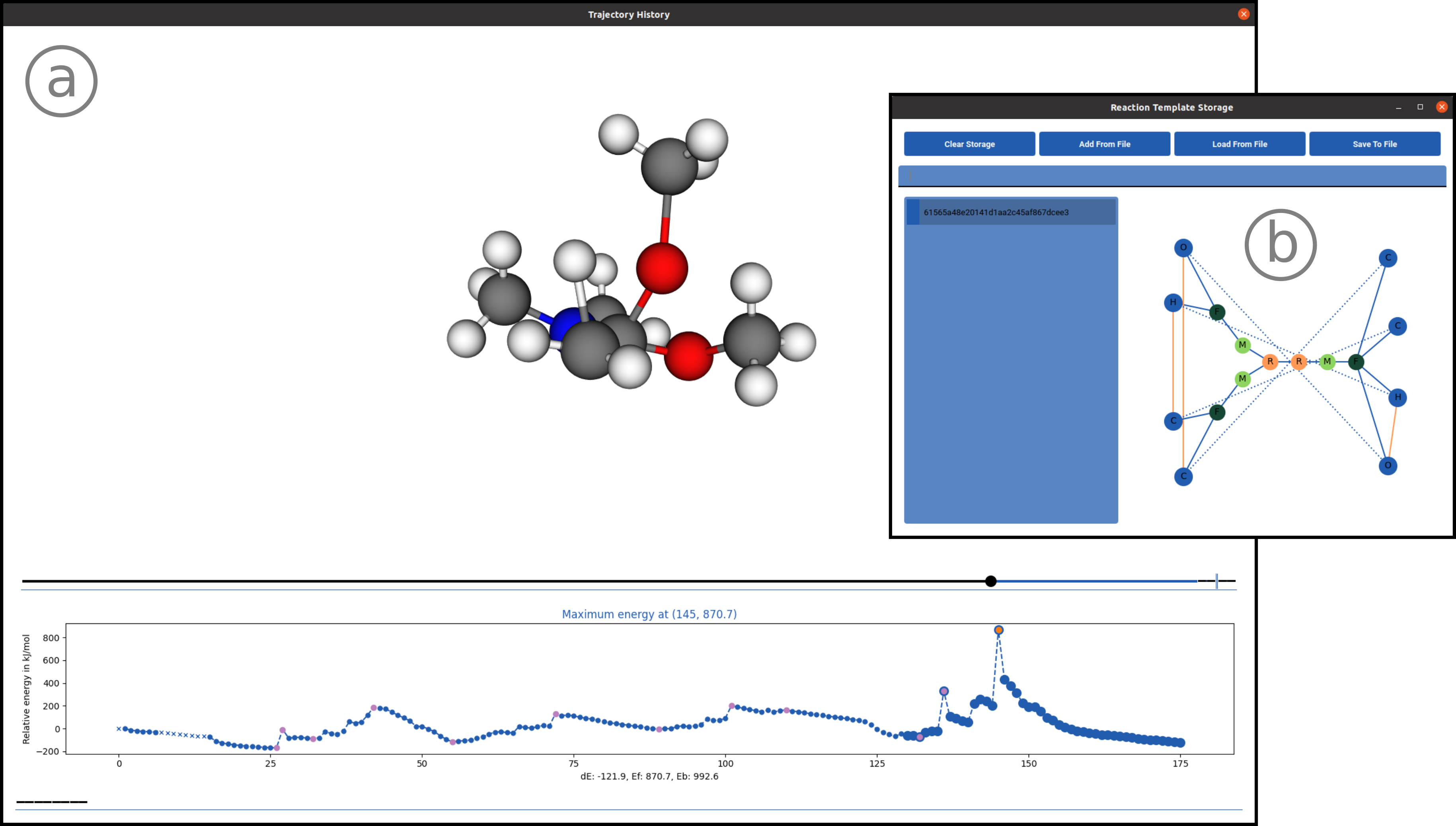}
    \caption{
    View of an interactively explored trajectory of the initial methanol dissociation reaction
    (a) and the reaction template storage dialogue showing the representation of a reaction template as defined in Ref.~\citenum{Unsleber2023a} (b).}
    \label{fig:templates}
\end{figure}

We now describe the visual tab for one special type of filtering, the possibility of defining so-called reaction templates.~\cite{Unsleber2023a}
These templates encode reactions based on the changes in the molecular graphs.
\textsc{Heron} allows for the extraction of reaction templates from any interactively explored trajectory of an elementary step and can load and save sets of reaction templates.
In the case of longer interactive exploration sessions, the operator is presented with a history of all modifications, can delete intermediate parts of this history, and picks start and end points, similarly to standard video editing software.
The view of a user-generated or loaded reaction trajectory and the template storage, including template preview, are shown in Figure~\ref{fig:templates}.
This feature is a prime example of how \textsc{Heron} facilitates the interoperability between different \textsc{Scine} packages and their dedicated features.
In this case, the interactive exploration, then data abstraction \textit{via} reaction templates, and its subsequent application within new automated reaction network explorations.

\subsubsection{Data Management and Statistics}
\label{subsec:database_viewer}
\begin{figure}[H]
    \centering
    \includegraphics[width=\textwidth]{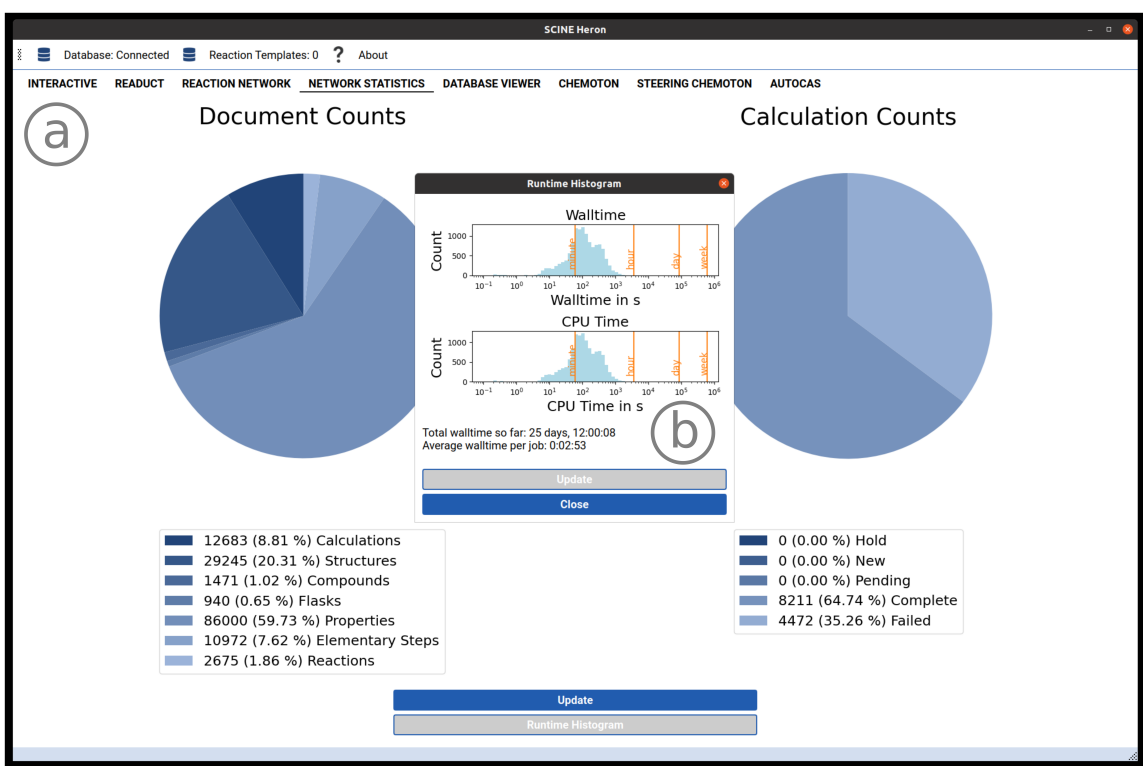}
    \caption{(a) 
    Basic database statistics in \textsc{Heron} and (b) dedicated on-demand view of runtime statistics.}
    \label{fig:db_stats}
\end{figure}

\textsc{Heron} has built-in features to interact with the reaction networks explored. The networks are maintained within the \textsc{Scine} \textsc{Database} schema~\cite{Unsleber2022, Database2023}.
\textsc{Heron} allows for browsing such databases as well as monitoring the exploration progress.
It is possible to view basic statistics, such as the number of calculations stored in the database according to their status (such as completed, pending, failed) in a statistics tab.
Further metrics, such as a histogram of runtimes per calculation, which requires more involved database queries, can be generated on demand.
Both features are shown in Figure~\ref{fig:db_stats}.

\begin{figure}[H]
    \centering
    \includegraphics[width=\textwidth]{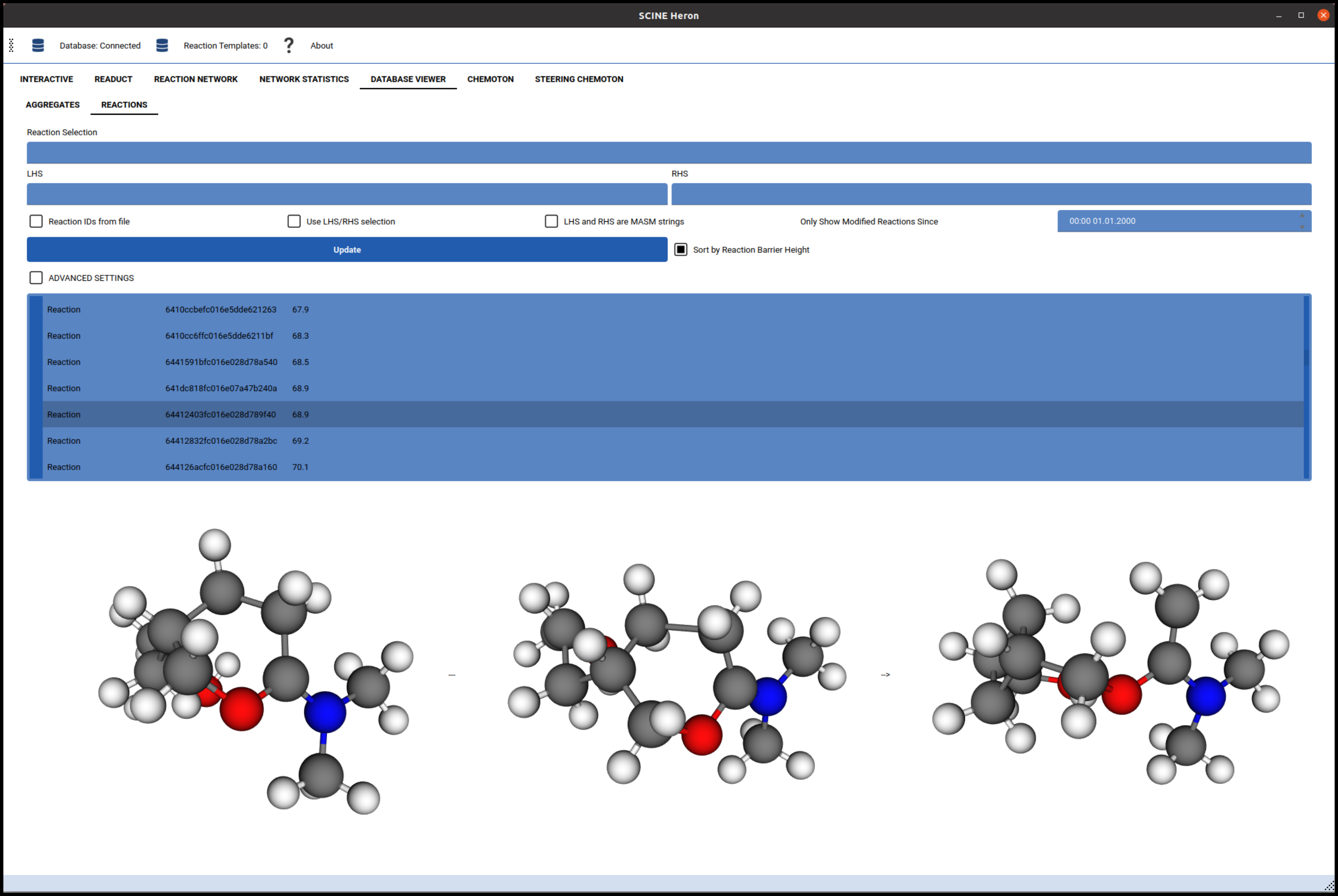}
    \caption{Table of all reactions in the Eschenmoser--Claisen network, with one single elementary step selected and displayed at the bottom.}
    \label{fig:db_browser}
\end{figure}

In addition to these basic network statistics, it is possible to assemble lists of reactions and compounds.
Since compounds comprise multiple structures (conformers) and, accordingly, reactions aggregate multiple elementary steps~\cite{Unsleber2020}, these entries are presented in a tree-like table.
Interactive three-dimensional renderings of the molecules can be viewed for compounds and structures.
For reactions and elementary steps, interactive three-dimensional renderings of all reactants, the transition state (if existent), 
and the products are displayed (see, for example, Figure~\ref{fig:db_browser}).
It is possible to filter these tables by graph comparisons (based on the \textsc{Scine}~\cite{Scine} software module \textsc{Molassembler}~\cite{Sobez2020, Molassembler2023}) and unique database identifiers.
Furthermore, the tables can be sorted by electronic energy.

\subsubsection{Network Visualization}
\label{subsec:crn}
An obstacle of understanding chemical reaction networks is their sheer size and the imbalanced and high degree of connectivity between the nodes.
Reaction networks can quickly grow to many thousand reactions connecting the associated chemical compounds.
Depending on the representation of a reaction network, each reaction and compound can be encoded as one or more nodes in a graph~\cite{Unsleber2020, Turtscher2023}.
In many cases, algorithms that optimize the graph layout by a pairwise repulsive potential between nodes fail to create an accessible two-dimensional representation of the network's complete graph~\cite{Steiner2022}.
Therefore, \textsc{Heron} avoids this type of representation.
Instead, \textsc{Heron}'s main network view focuses on smaller local regions of a network centering on one compound. It is possible to traverse the network from one such sub-graph to the next.
The direct surrounding of the central compound is always displayed.
Energy criteria can be set to filter and simplify the view of the direct surrounding further.
Moving the view to an adjacent compound (connected to the previous one by one reaction) is possible by clicking the computer mouse.
Besides this simple click-through mode, it is possible to jump to compounds directly if their unique identifier is known.
Finally, information can be displayed for the center compound or any other focused node, such as the three-dimensional structure of a single conformer or a single interpolated minimum energy path for the nodes representing reactions.

\begin{figure}[H]
    \centering
    \includegraphics[width=0.9\textwidth]{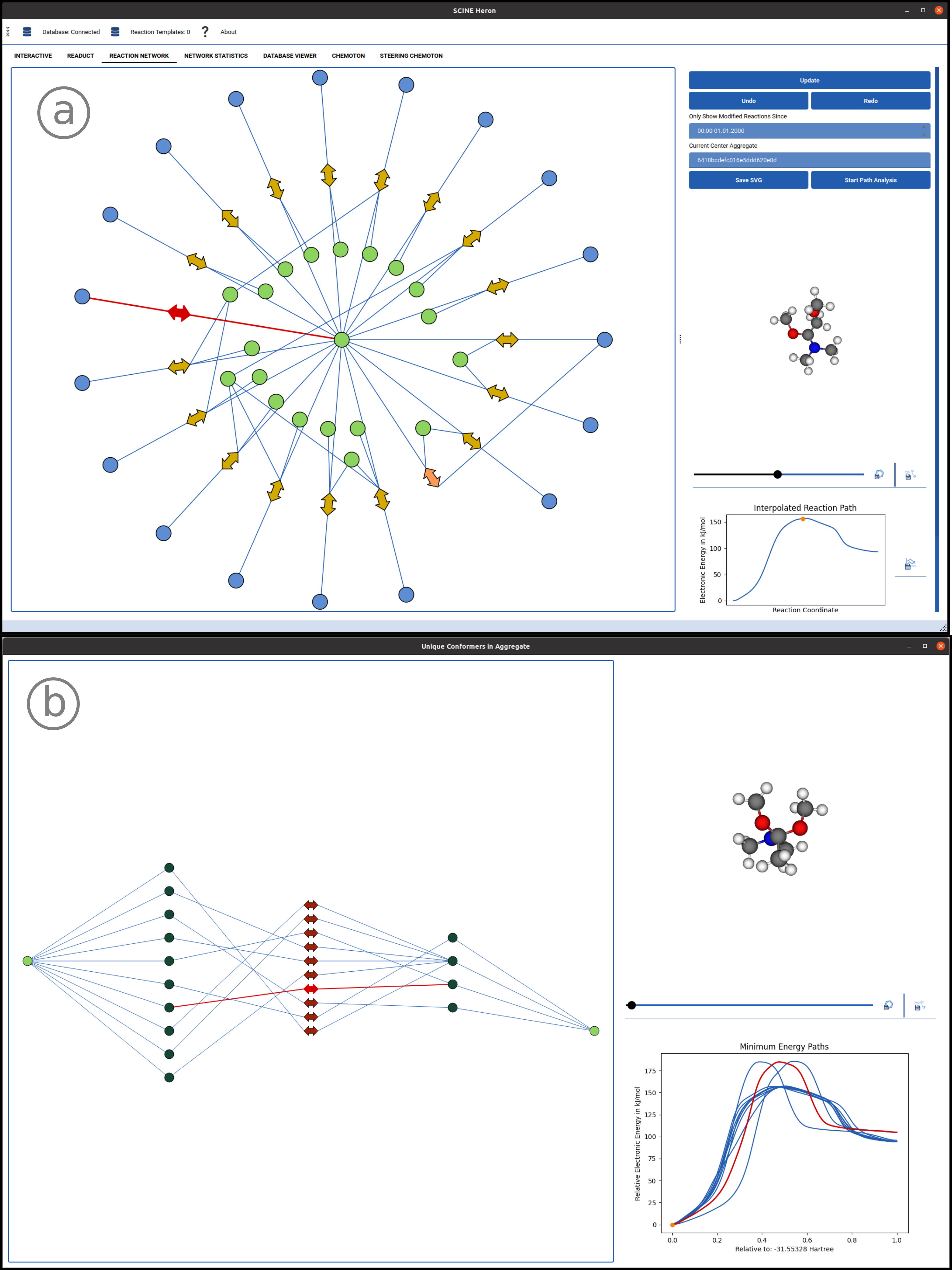}
    \caption{Basic Network representations in \textsc{Heron}.
    (a) Initial view of the network representing the Eschenmoser--Claisen rearrangement. Green nodes represent compounds, blue nodes represent flasks, light yellow arrows represent barrierless reactions, and orange arrows represent reactions with a transition state.
    (b) Expanded view of all elementary steps that were sorted into the reaction selected in (a). Elementary steps are sorted into the same reaction if they transform structures belonging to identical compounds or flasks. Dark blue nodes represent structures and red arrows represent elementary steps.}
    \label{fig:network_main}
\end{figure}

An example view taken from the Eschenmoser--Claisen reaction network is shown in Figure~\ref{fig:network_main}(a).
Right clicks on any node will produce an additional menu that allows for sending data contained in this (compound or reaction) node
to other parts of \textsc{Heron} or for generating additional views of the data.
An example of one such expansion view is shown in Figure~\ref{fig:network_main}(b), where all reported elementary steps aggregated within a single reaction node shown in Figure~\ref{fig:network_main}(a) are displayed.

\subsection{Reaction Path Identification}
\label{sec:pathfinder}

To facilitate network analyses, an operator typically would like to identify
which reactions lead to the transformation from the starting compound to the target compound.
Our path-finding algorithm \textsc{Pathfinder}~\cite{Turtscher2023} considers kinetic and stoichiometric constraints.
\textsc{Heron} then visualizes the most likely routes involving the most accessible reactions under given starting conditions.

\begin{figure}[H]
    \centering
    \includegraphics[width=\textwidth]{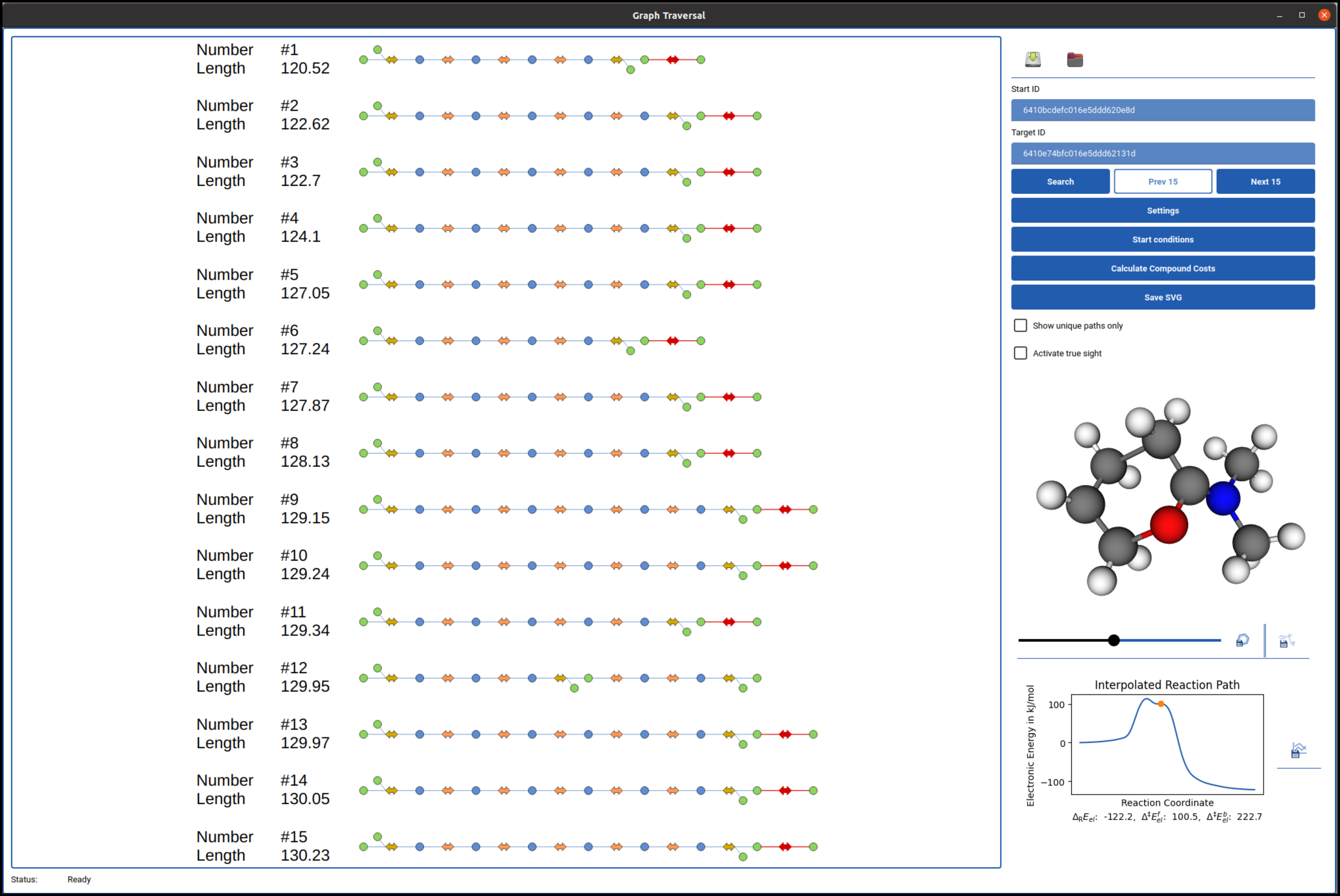}
    \caption{
    On the left part of the window, the path representation in \textsc{Heron} based on a \textsc{Pathfinder} analysis of a graph encoding kinetic and stoichiometric information.
    The shortest 15 paths connecting N,N-dimethylacetamide dimethyl acetal to the product of the Eschenmoser--Claisen rearrangement are shown. The highlighted reaction in red is the rearrangement reaction. 
    For each path, its rank is indicated to the right of the label 'Number', its length to the right of the label `Length'.
    The length is the sum of the lengths of the edges in each path.
    On the right, display options can be specified, specific compounds and reactions can be visualized, the latter including structural changes and energy profiles.}
    \label{fig:pathfinder}
\end{figure}

The theory behind encoding kinetic and stoichiometric information in a graph representation of chemical reactions with \textsc{Pathfinder} has been detailed in Ref.~\citenum{Turtscher2023}.
In \textsc{Heron}, the \textsc{Pathfinder} interface is accessed from the basic network representation (see section~\ref{subsec:crn}).
A new path analysis window
appears, and the operator can enter the IDs of the starting and target compounds in pre-defined fields
and adapt the settings of \textsc{Pathfinder} as well.
Different types of graphs can be constructed on the fly according to the setting of the operator.
One type is the basic graph.
In such a graph, all edges from a compound node to a reaction node have the length of 1 and are therefore identical.
All edges in the inverse direction, from a reaction node to a compound node, have a length of 0.
Hence, in such a graph the length of a path correlates to the number of reactions in this path.
Another graph type, the barrier graph, encodes barrier information in edges from a compound node to a reaction node.
The higher the barrier of a reaction, the longer the edge from the compound node to this reaction node.
The graphs built can be exported and imported.
Searching for connections triggers the construction of the underlying graph, basic or barrier,
and results in visualizing the first 15 shortest paths between the two nodes of interest, as depicted in Figure~\ref{fig:pathfinder}.

The paths are ordered from top to bottom according to increasing length.
The nodes of the paths are represented in the same style as in section~\ref{subsec:crn}.
Each path starts with the starting node (reactant) on the left and ends at the target node (product) on the right (see Figure~\ref{fig:pathfinder}).
Between these two end points, reaction and compound nodes occurring along the path are shown (see Figure~\ref{fig:pathfinder}, middle).
The compound nodes of incoming reagents required for a specific reaction are shown above the path, to the left of the affected reaction.
The compound nodes of outgoing side products of a specific reaction are shown below the path, to the right of the affected reaction.

In contrast to the network view, nodes (compounds as well as reactions) can occur more than once due to two reasons: 
First, one reaction or compound may be part of several distinct paths. 
Second, a compound may occur more than once in a single path since it can be a required reagent or formed as a by-product. 
To visualize this, upon hovering over one node, all identical nodes
are highlighted simultaneously in red, as indicated in Figure~\ref{fig:pathfinder}.
This allows for quick qualitative comparisons between paths and easily points to differences in competing reaction paths. 
Upon clicking on the 'Next 15' button on the right side of the window, more paths can be visualized by scrolling to the next 15 paths.

\subsection{Microkinetic Modeling}
\label{sec:kinetic_modelling}
The \textsc{Scine} software suite~\cite{Scine} provides microkinetic analysis tools through the program \textsc{KiNetX}~\cite{Proppe2018, Bensberg2023h} and an interface to the Reaction Mechanism Simulator~\cite{Liu2021, Johnson2023}.
Information on final and maximum concentrations and on concentration fluxes through a reaction network, are visualized in the representation of the chemical reaction networks.

\begin{figure}[H]
    \centering
    \includegraphics[width=\textwidth]{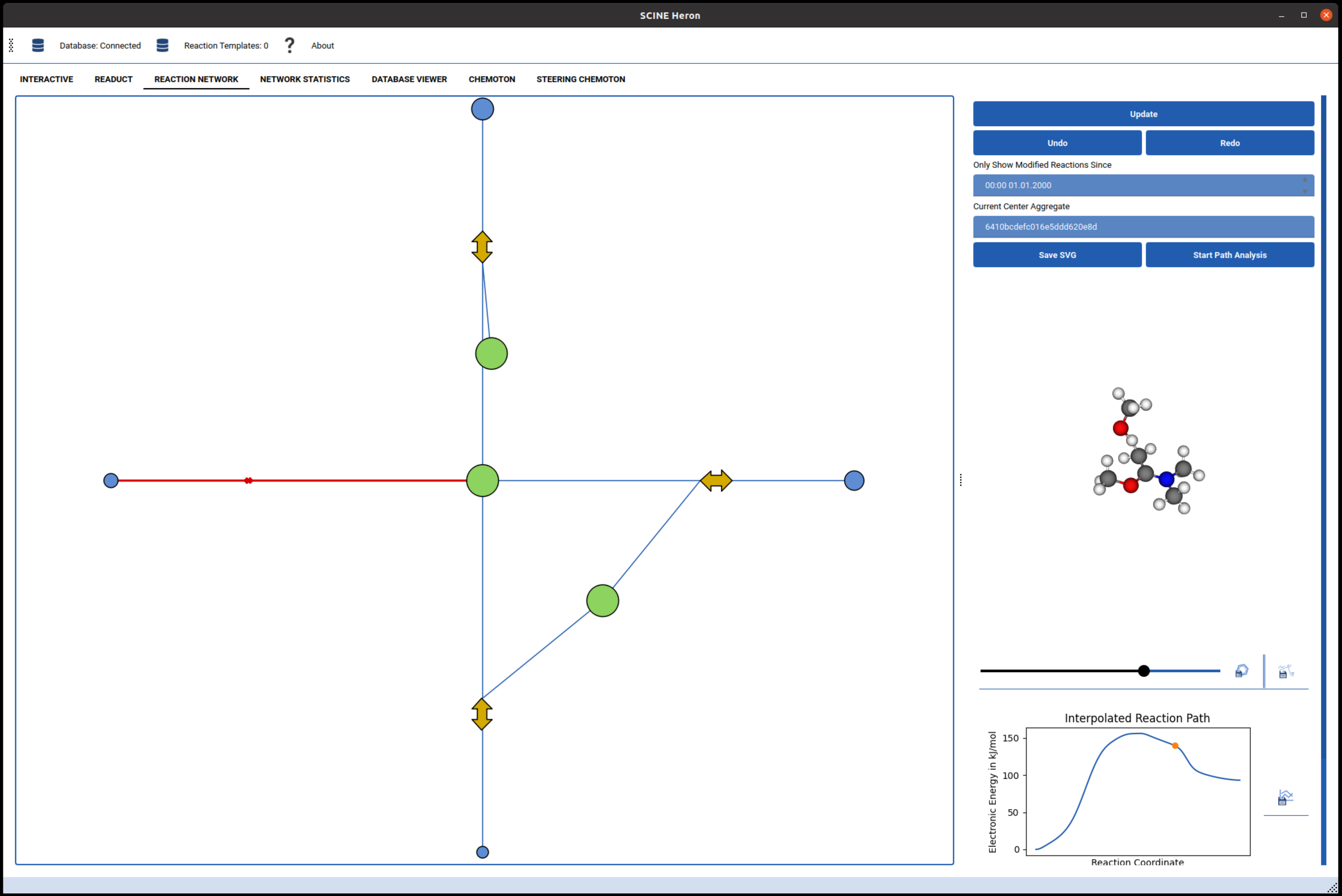}
    \caption{Reaction network view centered on N,N-dimethylacetamide dimethyl acetal of the Eschenmoser--Claisen network. The methanol dissociation was selected with the computer mouse and is highlighted in red. The spline interpolation for the electronic energy along the reaction's first elementary step and the reaction product are shown on the right. Only reactions $I$ are shown with an edge flux $F_I \geq 10^{-12}$. Additionally, the size of the arrows representing reactions are scaled by the corresponding edge flux $F_I$, and the circles representing compounds or flasks are scaled by their maximum concentration $c_i^\mathrm{max}$.}
    \label{fig:concentration_filtering}
\end{figure}

Reactions shown in the `interactive reaction network explorations' view can be filtered by their 
integrated concentration fluxes $F_I$ along a reaction $I$~\cite{Bensberg2023a}. This filtering approach eliminates a large number of reactions from the visualization that were found during the exploration but turned out to be kinetically irrelevant because of high reaction barriers or inaccessible reactants, compare Figure~\ref{fig:network_main}(a) and  Figure~\ref{fig:concentration_filtering}. 
In Figure~\ref{fig:network_main}(a), we focused the reaction network view on the Eschenmoser--Claisen starting reactant N,N-dimethylacetamide dimethyl acetal and omitted reactions with an GFN2-xTB~\cite{Bannwarth2020} reaction barrier larger than $262.5~\si{kJ.mol^{-1}}$, which retained $18$ reactions. If we filter with the condition $F_I \leq 10^{-12}~\si{mol.L^{-1}}$, as in Figure~\ref{fig:concentration_filtering}, only four reactions will be retained, of which the reaction highlighted in red is the methanol elimination step discussed in the mechanism.

The visualization in Figure~\ref{fig:concentration_filtering} also shows different radii of the circles representing compounds and flasks in the network by their maximum concentration $c_\mathrm{max}^n$ as well as widths of the arrows representing reactions by $F_I$, as introduced in Ref.~\citenum{Bensberg2023a}.
By scaling the size of the graphical elements in the reaction network, important compounds, flasks, and reactions are highlighted. This highlighting allows for a targeted exploration of the reaction network data through the interactive network representation.

\subsection{Data Refinement by Correlated Calculations with Automated Active Space Selection}
\label{sec:autocas}
Many electronic structures, in particular those of certain transition-metal complexes, transition states, bond-breaking processes, or excited states, usually require an accurate description of both static and dynamic electron correlation.
While dynamic correlation can be recovered in a black-box manner by coupled cluster theory~\cite{Bartlett2007}, static correlation emerges from determinants in the wave function expansion with a large expansion coefficient~\cite{Loewdin1955,Benavides2017}.
To describe static correlation, a complete active space (CAS~\cite{roos1980, roos1980b,Ruedenberg1982})
is typically chosen, including all orbitals required to build the determinants with a large weight.

However, choosing this active space is a non-trivial task, which had prevented applying active space-based methods in a black-box way for a long time.
We solved this problem with our \textsc{autoCAS} algorithm and the accompanying software package~\cite{Stein2016, Stein2016b, Stein2017, Stein2019, Stein2017b}. We exploit concepts from quantum information theory~\cite{Legeza2003, Rissler2006} to extract orbital information from an approximate CAS-type wave function to select orbitals for an active space.
This allows us to guarantee reproducibility, streamline active space calculations, and identify multi-configurational cases on a large scale~\cite{Unsleber2023}.

\textsc{Heron} replaces the initial graphical user interface of \textsc{autoCAS} v1.0.0~\cite{Stein2019}, which provided interfaces to \textsc{OpenMolcas}~\cite{Molcas2020, Molcas2023}
in combination with our DMRG program \textsc{QCMaquis}~\cite{Keller2015, Keller2016}.
The \textsc{Heron} interface is based on a reimplementation of \textsc{autoCAS}~\cite{Bensberg2023c} as a modular Python program.
As a result, it offers additional back-end options, currently \textsc{ChronusQ}~\cite{Williams-Young2019, Li2020} and \textsc{PySCF}~\cite{Sun2018, Sun2020} can be used in combination with \textsc{QCMaquis}.

The \textsc{autoCAS} tab of \textsc{Heron} allows the operator to run the \textsc{autoCAS} workflow~\cite{Stein2016} either fully automatically or step-wise, consisting of (i) the generation of initial orbitals through either Hartree--Fock or CAS self consistent field orbital optimization, (ii) an initial small-scale, unconverged DMRG calculation for the valence orbital space to identify the strongly correlated orbitals for the active space, and (iii) a final, fully converged CAS-type calculation including perturbation theory~\cite{Stein2016b}.
At every step, one could manually change settings, add or remove orbitals from the current active space (although this is usually not necessary and also not recommended), and activate the large active space protocol to approximate the active space search for valence spaces with more than 100 spatial orbitals~\cite{Stein2019}.
An orbital entanglement diagram can also be generated to characterize and further analyze the active space~\cite{Legeza2003, Rissler2006, Boguslawski2012, Boguslawski2013}. 

\begin{figure}[H]
    \centering
    \includegraphics[width=\textwidth]{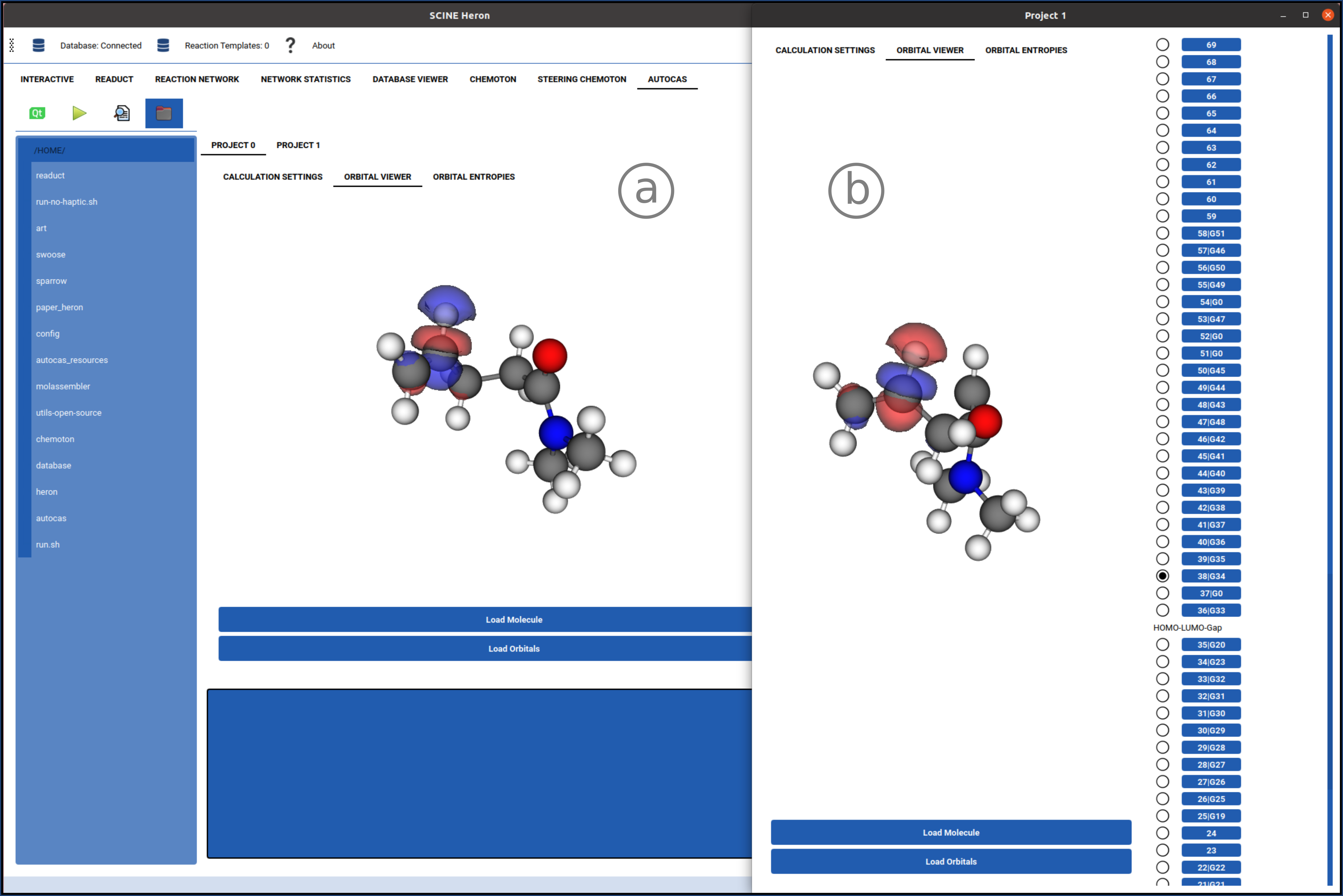}
    \caption{Illustration of the corresponding orbitals view in the \textsc{autoCAS} tab,  corresponding carbon--carbon $\sigma$ bond orbital of reactant and product of the Claisen rearrangement step in Figure~\ref{fig:eschenmoser_claisen_mechanism}.
    For each molecule, a tab is created in the left window (a). Each tab can be expanded in an additional window (b) to compare orbitals along reaction paths.}
    \label{fig:corresponding_orbitals}
\end{figure}

When calculating reaction energies and barriers with active space methods, the active orbital space should be consistent along the reaction path to ensure that corresponding orbitals are
always correlated for the reactants, the transition state, and the products.
Therefore, we combined the direct orbital selection mapping procedure~\cite{Bensberg2019a} implemented in the quantum chemistry program \textsc{Serenity}~\cite{Serenity2018, Niemeyer2022} with \textsc{autoCAS} to provide corresponding active orbital spaces along reaction paths~\cite{Bensberg2023b}.
In \textsc{Heron},
it is possible to load the orbital mapping into \textsc{Heron} to compare orbitals between structures. This orbital comparison is illustrated in Figure~\ref{fig:corresponding_orbitals}. In the right window in Figure~\ref{fig:corresponding_orbitals}, the \textsc{autoCAS} tab is shown with subtabs for each \textsc{autoCAS} project corresponding to molecular structures along the reaction path. In the left window of Figure~\ref{fig:corresponding_orbitals}, one of the projects was expanded to facilitate a comparison of orbitals along the reaction path. 
The example shows a corresponding carbon--carbon $\sigma$ bond orbital of reactant (left) and product (right) from the Claisen rearrangement step of the Eschenmoser--Claisen reaction in Figure~\ref{fig:eschenmoser_claisen_mechanism}. The orbitals can be selected for the active orbital space from the orbital viewer.

\subsection{Interactive Hybrid Models for Nanoscopic Systems}
\label{sec:interactive_qmmm}
The exploration of chemical structure and reaction space either by a real-time approach described in section \ref{sec:interactive} or by automated algorithms in \textsc{Chemoton}~\cite{Unsleber2022, Chemoton2023} in section~\ref{subsec:exploration} becomes computationally challenging for large nanoscale systems such as biomolecules and support-based catalysts.
Multiscale models, such as quantum mechanics/molecular mechanics (QM/MM) hybrid models, allow one to hold a ``quantum magnifying glass''\cite{Csizi2023} over a local region within a nanoscopic system. In this way, the electronic structure seen in the magnifying glass can be exploited for general reactivity analysis. Application of these models in the context of automated exploratory strategies is hampered by complicated model construction, which we alleviated by automatizing all steps in a single software framework, called \textsc{Swoose}~\cite{Swoose2021}, which is a \textsc{Scine} module for multiscale modeling.
\textsc{Heron} implements an interface to \textsc{Swoose}.

Within \textsc{Heron}, the full QM/MM hybrid model can be built interactively from scratch~\cite{Csizi2023}. In the first step, a system-focused atomistic model (SFAM)~\cite{Brunken2020} can be parametrized from quantum mechanical reference data, for which
only the molecular structure and a text file storing information about charged groups must be provided. 
If this shall be accomplished, in real time, ultra-fast reference data generation for MM model parametrization will be ensured by applying semiempirical electronic structure methods~\cite{Bosia2022, Bosia2023, Sparrow2023}. However, any electronic structure method available in the \textsc{Scine} infrastructure can be employed as well, but with increased timings for the reference calculations.
Additionally, reference data generation employing a fragmentation algorithm, which requires a database connection, is incorporated in \textsc{Heron}, allowing one to parameterize an SFAM model for nanoscopic systems in a reasonable time frame.
As an alternative to SFAM, the general Amber force field~\cite{Wang2004} can be applied for the MM calculation.

In the next step, the corresponding quantum region can be selected automatically within \textsc{Heron} (i) by the QM/SFAM formalism developed in our group~\cite{Brunken2021} with reference calculations carried out in the background with respect to which an error is optimized, (ii) by a radial distance cut-off around a selected central atom,
or (iii) by manual selection of the corresponding atoms. 
The SFAM and QM/SFAM settings can be adjusted within \textsc{Heron}, with default settings optimized for interactive explorations.

\begin{figure}[H]
    \centering
    \includegraphics[width=1.0\textwidth]{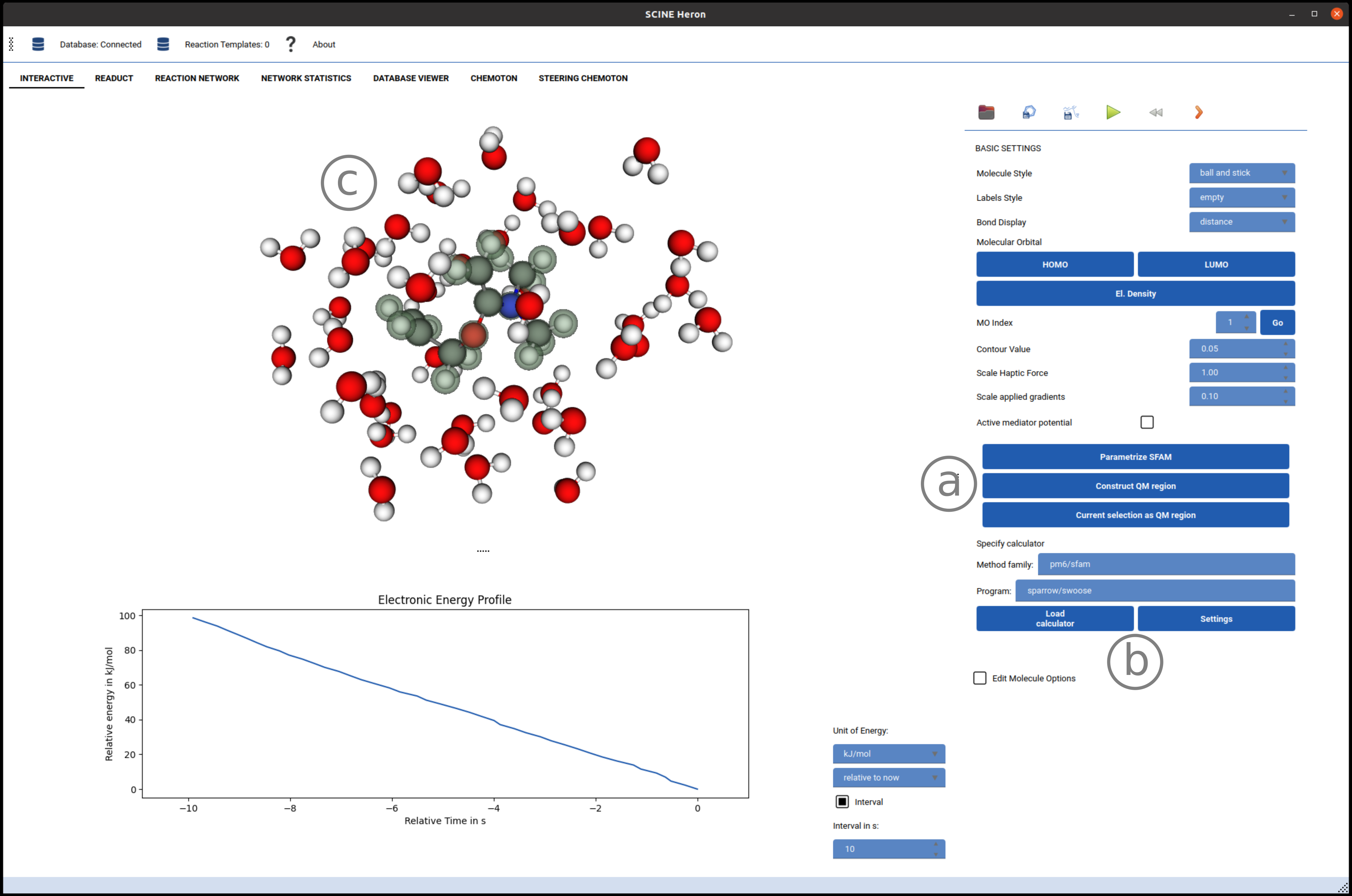}
    \caption{The interactive module with QM/MM functionality. The buttons for SFAM parametrization and QM/SFAM region selection are shown on the right (a) and will automatically appear if the \textsc{Swoose} program is available. The quantum and classical methods and the corresponding programs can be given in the 'Specify Calculator' interface (b), separated by a slash (for instance: PM6/SFAM, and  \textsc{Sparrow/Swoose}). The quantum region is highlighted through green spheres (c) in the molecular viewer.}
    \label{fig:interactive-qmmm}
\end{figure}

The resulting hybrid model can be subjected to automated optimization routines provided by the software module \textsc{ReaDuct} or explored within the interactive framework.
Reaction paths can be explored with \textsc{Chemoton} with multiscale models by transplantation of model structures into the MM scaffold and subsequent assessment of environment effects on reaction steps found in the core region~\cite{Csizi2023}.
Large solvated complexes can be described by an MM solvent and a quantum mechanically described solute.
In Figure \ref{fig:interactive-qmmm}, compound \textbf{4} of the mechanism in Figure~\ref{fig:eschenmoser_claisen_mechanism}(b) is explicitly solvated in water by employing the microsolvation protocol presented in Ref.~\citenum{Simm2020}. The selected quantum region is marked by green spheres.

\section{Conclusions}
\label{sec:outlook}
In this work, we presented \textsc{Heron}, a new graphical user interface for exploring reactive chemical systems within the \textsc{Scine} software framework.
\textsc{Heron} offers (i) an interface for real-time (haptic) studies of molecules and their reactions, (ii) the possibility to carry out structure optimizations and similar molecular structure-based algorithms, (iii) a platform to analyze chemical reaction networks, (iv) interlaced microkinetic modelling, and (v) refinement by correlated calculations including automated active space selection with subsequent multi-configurational calculations.
All these features are linked to the automated exploration of reaction networks, the main focus of \textsc{Heron}.
This unique combination of features makes \textsc{Heron} a premier platform for studying chemical processes for non-expert and expert users.
\textsc{Heron} is available open source and free of charge~\cite{Heron2022}.
A key advantage of \textsc{Heron} is the interoperability of all features listed in this work. While each feature on its own provides advanced capabilities for studying chemical reaction networks or the underlying chemistry of molecules, it is the possibility of seamless data transfer between each task that will allow for new ways of interacting with quantum chemical data.

\section*{Acknowledgments}
\label{sec:acknowledgments}

This publication was created as part of NCCR Catalysis, a National Centre of Competence in Research funded by the Swiss National Science Foundation (grant number 180544).
MS gratefully acknowledges a Swiss Government Excellence Scholarship for Foreign Scholars and Artists. Financial support is also gratefully acknowledged by the ETH+ Future Learning Initiative, ETH grants ETH-43 20-2 and ETH-44 20-1, as well as an ETH Innovedum project. We thank the Scientific IT Services of ETH Zurich and eXact lab for programming support.

\section*{Data Availability Statement}
The kinetically driven reaction network is provided as part of a larger data set on Zenodo~\cite{kinetic_dataset2024}.
All data newly generated in this work is provided in a separate Zenodo repository~\cite{HeronZenodoSI2024}.

\section*{Software Availability Statement}
\textsc{Scine Heron} as well as all \textsc{Scine} modules and third-party software mentioned in this manuscript are available open source via the links provided below. All links were accessed on 2024-07-19.
\begin{itemize}
    \item \textsc{Scine} modules
    \begin{itemize}
        \item \textsc{Art}: \url{https://github.com/qcscine/art} %\url{}
        \item \textsc{autoCAS}: \url{https://github.com/qcscine/autocas} %\url{https://zenodo.org/record/7620381}
        \item \textsc{Chemoton}: \url{https://github.com/qcscine/chemoton} %\url{https://zenodo.org/record/7928104}
        \item \textsc{Database}: \url{https://github.com/qcscine/database} %\url{https://zenodo.org/record/7928096}
        \item \textsc{Heron}: \url{https://github.com/qcscine/heron} %\url{https://zenodo.org/record/7038388} 
        \item \textsc{KiNetX}: \url{https://github.com/qcscine/kinetx} %\url{https://zenodo.org/records/7928112}
        \item \textsc{Molassembler}: \url{https://github.com/qcscine/molassembler} %\url{https://zenodo.org/record/7928074}
        \item \textsc{QCMaquis}: \url{https://github.com/qcscine/qcmaquis} %\url{https://zenodo.org/records/10843173}
        \item \textsc{ReaDuct}: \url{https://github.com/qcscine/readuct} %\url{https://zenodo.org/record/7928089}
        \item \textsc{Sparrow}: \url{https://github.com/qcscine/sparrow} %\url{https://zenodo.org/record/7928079}
        \item \textsc{Swoose}: \url{https://github.com/qcscine/swoose} %\url{https://zenodo.org/record/5782877}
    \end{itemize}
    \item Third-party software
    \begin{itemize}
        \item \textsc{ChronusQ}: \url{https://github.com/xsligroup/chronusq_public}
        \item \textsc{OpenHaptics}: \url{https://support.3dsystems.com/s/article/OpenHaptics-for-Linux-Developer-Edition-v34?language=en_US}
        \item \textsc{Serenity}: \url{https://github.com/qcserenity/serenity}
        \item \textsc{xtb}: \url{https://github.com/grimme-lab/xtb}
    \end{itemize}
\end{itemize}

\appendix
\section{Technical Details}
\label{app:design}
\textsc{Heron} has been primarily written in the Python programming language. Being a widely established, general-purpose programming language, it features
both long-term stability and flexibility which are required to maintain \textsc{Heron} over a long time and continuously adapt it to future needs. 
Furthermore, code development with Python is often faster than with compiled languages such as C++~\cite{Prechelt2000}.
Moreover, installing a Python package is trivial on most computer systems. 
However, a well-known drawback of an interpreted language such as Python is that it is rather slow compared to compiled languages such as C++ and Fortran. 
Fortunately, for most aspects of \textsc{Heron}, Python code runs fast enough on standard contemporary computer hardware.
Moreover, Python provides established and reliable mechanisms to offload compute-intensive tasks to very efficient code written, for example, in C++. 
Many dependencies of \textsc{Heron} take advantage of this; for example, the widely-used Python package NumPy~\cite{Harris2020} for numerical computations relies on C functions for fast linear algebra operations. 
Also, \textsc{Heron} contains C++ code for interfacing haptic devices.

\textsc{Heron} relies on several third-party libraries. The most important ones are PySide2~\cite{pyside2} (interfacing the well-known
Qt software~\cite{qt} for creating graphical user interfaces), VTK~\cite{Hanwell2015} (a toolkit for three-dimensional computer graphics), Matplotlib~\cite{Hunter2007}
(which provides functionality for visualizations), SciPy~\cite{Virtanen2020} (a library containing many routines for scientific computing), and NumPy~\cite{Harris2020}.

The graphical user interface itself aims to be as intuitive as possible.
Each tab or window is divided into a primary display and a settings area. 
Within the latter, key options and settings are always shown, whereas expert options are hidden by default. 
The individual areas can be adjusted in size (and even completely collapsed) as desired by the user. 
By default, two different color schemes are supplied.
First, a ``dark theme'' featuring a dark background and light foreground colors, and second a ``light theme'' with a bright background and dark foreground colors are included.
Simple XML-files define coloring schemes, so creating custom color schemes and expanding on existing ones is straightforward.

\textsc{Heron} is distributed under the permissive 3-clause (``new'' or ``modified'') BSD license. This permits use and redistribution, both
with and without modification, by virtually anyone. As a result, there are no legal hurdles to consider before adopting \textsc{Heron} and incorporating
it into existing workflows.

\section{Exploration Protocol for the Eschenmoser--Claisen Rearrangement\label{sec:exploration_details}}

The reaction network is explored step-wise in the KIEA algorithm. First, the algorithm searches for all possible reactions of the educts, and then the concentration flux through the discovered reaction network is simulated by micro-kinetic modeling. KIEA then decides which compounds should be probed for reactions based on the concentration fluxes calculated by the micro-kinetic modeling. This procedure of reaction network exploration, micro-kinetic modeling, and selection of new compounds is continued until no further compounds are trialed for reactions.

Searching for chemical reactions is handled by \textsc{Chemoton}. \textsc{Chemoton} automatically probes compounds for chemical reactions with first-principles calculations, as described in Ref.~\cite{Unsleber2022}. For these calculations, we modeled the electronic structure with the semi-empirical tight-binding model known as GFN2-xTB\cite{Bannwarth2020} and the generalized Born and surface area implicit solvation model\cite{Onufriev2004, Sigalov2006} parameterized for GFN2-xTB\cite{Bannwarth2020} to model toluene as a solvent (referred to as GFN2-xTB(Toluene) in the following sections). Furthermore, we refined the electronic energies for all stationary points on the potential energy surface with density-functional theory (DFT), the exchange--correlation functional known as PBE0\cite{Adamo1999}, the def2-TZVP basis set\cite{Ahlrich2005}, the D3 dispersion correction\cite{Grimme2010a} with Becke--Johnson damping\cite{Grimme2011}, the conductor-like solvation model\cite{Klamt1993} to model toluene, and the quantum chemistry program \textsc{Turbomole}\cite{turbomole742}.
The concentration flux through the reaction network is modeled with the \textsc{Reaction Mechanism Simulator}(RMS) micro-kinetic modeling program\cite{Liu2021, Johnson2023}, assuming a constant temperature of $150~\si{\celsius}$, start concentrations for all reactants of $1~\si{mol.L^{-1}}$, and propagating for $10~\si{h}$. The reaction rate constants $k_I^{+/-}$ for the micro-kinetic modeling were calculated according to transition state theory\cite{Eyring1935, Truhlar1996} as
\begin{align}
    k_I^{+/-} = \Gamma \frac{k_BT}{h}\exp \left[-\frac{\Delta G^{\ddagger +/-}_I}{R T}\right]~,
\end{align}
where $+/-$ indicates the direction of the reaction $I$ (forward +, backward -), $\Gamma$ is the transmission coefficient, $k_B$ is Boltzmann's constant, $R$ is the molar gas constant, $T$ is the temperature, $h$ is Planck's constant, and $\Delta G^{\ddagger +/-}_I$ is the free energy of activation of $I$ and its direction at $150~\si{\celsius}$ and $1~\si{mol.L^{-1}}$. Furthermore, we assumed $\Gamma = 1$ and calculated the free energies from the DFT electronic energies and the commonly used rigid rotor, harmonic oscillator, and particle in a box approximation based on the GFN2-xTB structures. The harmonic vibrational frequencies were calculated with GFN2-xTB(Toluene). We will denote this combination of DFT electronic energies and GFN2-xTB(Toluene) free energy corrections as PBE0-D3BJ(Toluene)//GFN2-xTB(Toluene) in the following.

According to our concentration-flux-steered reaction network exploration algorithm\cite{Bensberg2023a}, we explored unimolecular reactions for all compounds $n$ with an integrated concentration flux $c_\mathrm{flux}^n$ of more than $1\cdot 10^{-3}~\si{mol.L^{-1}}$ and bimolecular reactions between compounds $n$ and $m$ if the product of their maximum concentrations $c_\mathrm{max}^n c_\mathrm{max}^m$ exceeded $1\cdot 10^{-2}~\si{mol^2.L^{-2}}$.

To guarantee a rapid exploration of the reaction network, we restricted the reaction coordinates probed for reactions according to chemical textbook knowledge of the reaction mechanism of the Eschenmoser--Claisen rearrangement, analogous to the approach in Ref.~\cite{Bensberg2023a}.
Hydrogen atoms were only probed for reactions if they were in a 2-position to an acetal, carbonyl, or imino group or if they were part of a carboxyl or amide group. Carbon atoms were considered in the reaction coordinates if they were part of a carbonyl or imino group, if they were in a 1-position to an acetal, imino, or carbonyl group, or if they were olefinic. Nitrogen atoms were considered reactive if they were part of an amine group, and oxygen atoms were considered reactive if they were part of a hydroxyl or carboxyl group.
Furthermore, we only probed reaction coordinates if the atoms involved were differently polarized, \emph{i.e.} if we assigned different polarization indicators to them. These indicators were assigned to two neighboring atoms according to their difference in Pauling electronegativity if this difference was larger than $0.4$. We chose a minimum difference of $0.4$ to avoid any assignment of polarization indicators to CH$_X~(X= 1,2,3)$ groups. To ensure that acidic protons were still included in the reaction trials, we assigned a positive polarization indicator. In addition, olefinic carbon atoms were assigned a positive and a negative polarization indicator.

The KIEA algorithm considered no new compounds for the automatic exploration after the sixth micro-kinetic modeling simulation.

%\bibliography{references}
\providecommand{\latin}[1]{#1}
\makeatletter
\providecommand{\doi}
  {\begingroup\let\do\@makeother\dospecials
  \catcode`\{=1 \catcode`\}=2 \doi@aux}
\providecommand{\doi@aux}[1]{\endgroup\texttt{#1}}
\makeatother
\providecommand*\mcitethebibliography{\thebibliography}
\csname @ifundefined\endcsname{endmcitethebibliography}
  {\let\endmcitethebibliography\endthebibliography}{}

\end{document}